\documentclass[aps,prl,twocolumn,superscriptaddress]{revtex4-1}
\usepackage{amsmath,epsfig}
\usepackage{graphicx}
\usepackage{color}
\def\be{\begin{equation}}
\def\ee{\end{equation}}
\def\bea{\begin{eqnarray}}
\def\eea{\end{eqnarray}}

\begin{document}
\title{Comment on ``A classical model for asymmetric sidebands in cavity optomechanical measurements''}

\author{Amir H. Safavi-Naeini} \affiliation{Thomas J. Watson, Sr., Laboratory of Applied Physics,
California Institute of Technology, Pasadena, CA 91125}
\email{amir.safavi@gmail.com, opainter@caltech.edu}
\author{Oskar Painter}
\affiliation{Thomas J. Watson, Sr., Laboratory of Applied Physics,
California Institute of Technology, Pasadena, CA 91125}

\date{\today}

\begin{abstract}
We respond to a recent manuscript by Tsang [arXiv:1306.2699], on whether the measurement presented in Safavi-Naeini et al. [Phys. Rev. Lett. 108, 033602 (2012)] can be explained ``without reference to quantum mechanics''. We show that the fully classical analysis provided by Tsang, and previously by Safavi-Naeini et al. [New J. Phys. 15, 035007 (2013)], has been ruled out by our published data. In addition, we discuss the role of the mathematical formulation used on the interpretation of the asymmetry effect, as has previously been considered by Khalili et al. [Phys. Rev. A 86, 033840 (2012)].
\end{abstract}
\maketitle

Recent experiments in optomechanics have begun to claim observation of quantum phenomena~\cite{Murch2008,Safavi-Naeini2012,Brahms2012,Brooks2012,Purdy2013,Safavi-Naeini2013c,Purdy2013a}. In many of these measurements, the basic experimental technique is noise spectroscopy of the light coming from an optomechanical system via correlation measurements and homo(hetero)dyne detection. A linear amplifier model, where the optomechanical system is considered to be a network capable of transfering and modifying classical and quantum noise~\cite{Botter2012}, can be used to interpret all of these results. Such a model can lead to a scattering matrix or input-output formulation, where the ``system'' variables, i.e., the optical and mechanical resonator are eliminated. Thus, all detected noise spectra can be understood by the linear relations between different input-output ``bath'' operators (optical, and mechanical) and their correlations~\cite{Khalili2012,Safavi-Naeini2013b}. Since the quantum noise is Gaussian, these correlations can be mimicked by a classical noise process. The best that an experimentalist using the continuous detection of light can do in this situation, is to make the classical noise scenario exceedingly unlikely. This has been convincingly achieved in the aforementioned experimental results~\cite{Murch2008,Safavi-Naeini2012,Brahms2012,Brooks2012,Purdy2013,Safavi-Naeini2013c,Purdy2013a}. In a recent manuscript by Tsang~\cite{Tsang2013}, several claims with respect to the result published in Ref.~\cite{Safavi-Naeini2012} are made. We address these claims here.

Tsang's first claim is that he has demonstrated ``a classical stochastic model, without any reference to quantum mechanics'' that can ``reproduce this asymmetry.'' Tsang assumes that there is classical noise present and shows that classical noise leads to an asymmetry. We have precisely, and in detail, studied this noise model in Ref.~\cite{Safavi-Naeini2013b}. In Ref.~\cite{Safavi-Naeini2013b} we go further and explore other experimental consequences of this hypothetical classical noise model. We rule out various types of classical noise by independently measuring the noise properties of the laser, and using an attenuator to vary the probe laser power used, finding that the observed asymmetry is unaffected. In particular, we find that the hypothetical classical noise required to give rise to the observed asymmetry needs to have magnitude and spectral properties exactly that of optical vacuum noise. Moreover, to explain our full experimental results, one would require this hypothetical classical noise to be unaffected by attenuation, spectral filtering, or changes in the laser power. Such classical background electromagnetic fluctuations would also have to be several orders of magnitude outside of thermal equilibrium with the surrounding environment. Also it would need to be present not only on the input channel of the optical waveguide, but also on the counter-propagating direction, as well as on all the intrinsic loss channels of the optical cavity comprised of absorption and scattering losses, over which we have no control. By ignoring the actual experimental data (presented in Refs.~\cite{Safavi-Naeini2012,Safavi-Naeini2013b}), Tsang does not address these  consequences of this simple classical noise model, and avoids presenting the exceedingly unlikely and ``conspiratorial'' properties of the classical noise that would be required to reproduce our full result. We believe that our experiments and follow-up analysis~\cite{Safavi-Naeini2012,Safavi-Naeini2013b} make such a fully classical explanation of our result exceedingly unlikely, which is the best that one can do with the currently demonstrated measurement techniques in optomechanics. It is also important to note that equally careful analyses of the classical noise properties are made in experiments which demonstrate no quantum effects, but make strong claims about the \textit{phonon occupation numbers} of the mechanical system~\cite{Groeblacher2009a,Park2009,Schliesser2009,Rocheleau2010,Teufel2011,Chan2011,Verhagen2012}, and Tsang's analysis applies equally to these works as well.

Tsang's second claim is predominantly about interpretation, and Ref.~\cite{Tsang2013} provides neither a novel nor a comprehensive interpretation of the detected spectra. Considering that classical noise may not be the source of the detected asymmetry, Tsang concedes that ``Quantum mechanics is needed to explain the asymmetry ...", then claims ``... but even then the limit is on the optical noise power $S_A$ and says nothing about the mechanical energy,'' a point considered in detail in Ref.~\cite{Khalili2012}. We believe this is a matter of interpretation of the underlying full quantum mechanical theory of the experiment. It is possible that there is a theory where the mechanical system is treated classically, and the optical system is treated quantum mechanically. We are not aware of such a theory, and were unable to arrive at a consistent formulation. In any case, this is not what has been presented by Tsang~\cite{Tsang2013}. As such we focus on the full quantum theory address the question as to whether the sideband asymmetry is a signature of the zero-point motion, or whether it is simply due to the bath variables. The interpretation used by Tsang~\cite{Tsang2013} is linked strongly to input-output formalism utilized. It is hardly surprising that by following a scattering matrix treatment, where the detected spectra are found by eliminating the \textit{system} operators and calculating the detected spectra only by referring to the \textit{bath} operators, one is led to an interpretation where only the fluctuations of the bath operators are important. Equivalently one can eliminate all operators but that of the mechanical resonator, as has been demonstrated systematically by Wilson-Rae et. al.~\cite{Wilson-Rae2007,Wilson-Rae2008}, to arrive at a master equation (dependent on laser detuning and power) for the mechanical subsystem. This is also the viewpoint taken by Marquardt et al.~\cite{Marquardt2007}, where the spectrum of the optical force on the mechanical system is calculated to find the phonon absorption and emission rates. It is in this context that it can be claimed that the asymmetry in the emission and absorption rates arises from the zero-point motion of the mechanical resonator, in direct analogy to the early experiments with trapped ions~\cite{Diedrich1989}. The two differences between the trapped ion experiments and Ref.~\cite{Safavi-Naeini2012} are that the long optical resonator lifetime means only one sideband is generated at a time, and that homodyne measurement is used as opposed to photon counting -- both of these difference are considered thoroughly in section 2.D of Ref.~\cite{Khalili2012}.

Finally we reiterate that as with Ref.~\cite{Safavi-Naeini2012,Brahms2012}, the recent experiments demonstrating quantum effects such as squeezed light~\cite{Brooks2012,Safavi-Naeini2013c,Purdy2013a} and radiation-pressure shot-noise (RPSN)~\cite{Purdy2013} all demonstrate in some form the modification of the quantum optical noise by the optomechanical system. In the squeezing experiments, this leads to sub-shot-noise fluctuations of the reflected light, while in the RPSN experiment independent cross-correlation measurements between the detected motion and the radiation-pressure back-action force~\cite{Heidmann1997,Borkje2010} are made.  In the works claiming to demonstrate the zero-point motion~\cite{Safavi-Naeini2012,Brahms2012}, the observed asymmetry in the motional sidebands can also be considered to arise from quantum back-action interference~\cite{Khalili2012,Safavi-Naeini2013b}. It is important to note however, that in contrast to the other experiments, only these measurements require the mechanical system to be close to its quantum ground-state in order to observe a signature -- both squeezing and RPSN heating have been detected with mechanical resonators in highly thermal states. In the measurements of zero-point motion, we consider both the sum, and the difference of the sideband amplitudes. The sum of the red and blue sideband amplitudes contains information about the mechanical noise and the total energy in the system, while it is the ratio of the sidebands, which is found to deviate from the classical theory. 

Looking forward, we believe it is of great scientific interest to demonstrate quantum phenomena such as violations of Bell's inequalities, generation non-Gaussian states, or observation of quantum jumps with optomechanics. The current, perhaps weaker demonstrations of non-classical effects~\cite{Murch2008,Safavi-Naeini2012,Brahms2012,Brooks2012,Purdy2013,Safavi-Naeini2013c,Purdy2013a}  can be viewed as a stepping-stone towards that goal.

The authors would like to thank K.\ Lenhert, D.\ Stamper-Kurn, H.\ Miao, K.\ Hammerer, and Y.\ Chen for insightful conversations. This work was supported by the DARPA/MTO ORCHID program through a grant from AFOSR, by the Institute for Quantum Information and Matter, an NSF Physics Frontiers Center with support of the Gordon and Betty Moore Foundation, and the Kavli Nanoscience Institute at Caltech. ASN acknowledges support from NSERC.


\bibliographystyle{apsrev}
\bibliography{Mirror}


\end{document}